\documentclass[11pt]{article}

\usepackage[preprint,nonatbib]{nips_gibs}

\usepackage[utf8]{inputenc} % allow utf-8 input
\usepackage[T1]{fontenc}    % use 8-bit T1 fonts
\usepackage{hyperref}       % hyperlinks
\usepackage{url}            % simple URL typesetting
\usepackage{booktabs}       % professional-quality tables
\usepackage{amsfonts}       % blackboard math symbols
\usepackage{nicefrac}       % compact symbols for 1/2, etc.
\usepackage{microtype}      % microtypography

\usepackage[english]{babel}

\usepackage{amsmath}
\usepackage{amssymb}
\usepackage{amsthm}

\usepackage{graphicx}

\usepackage{enumitem}
\setitemize{noitemsep,topsep=2pt,parsep=2pt,partopsep=2pt}
\setenumerate{noitemsep,topsep=2pt,parsep=2pt,partopsep=2pt}

\usepackage[parfill]{parskip}
\usepackage{tabularx}

\usepackage{subfig}
\newcommand{\DM} {\normalfont{\textsf{DM}}}
\newcommand{\dDM} {\normalfont{\textsf{dDM}}}
\newcommand{\dPDM} {\normalfont{\textsf{dPDM}}}

\newcommand{\fraB}{\ensuremath{\mathcal{B}}}
\newcommand{\fraN}{\ensuremath{\mathcal{N}}}
\newcommand{\fraS}{\ensuremath{\mathcal{S}}}
\newcommand{\fraT}{\ensuremath{\mathcal{T}}}

\newcommand{\Z}{\ensuremath{\mathbb{Z}}}

\theoremstyle{definition}

\begin{document}

\title{Secure Exchange of Digital Goods in a Decentralized Data Marketplace}

\author{Ariel Futoransky \\
	Wibson \& Disarmista
	\And Carlos Sarraute \\
	Wibson 
	\And Ariel Waissbein \\
	Wibson  \& Disarmista
	\AND Daniel Fernandez \\
	Wibson
	\And Matias Travizano \\
	Wibson 
	\And Martin Minnoni \\
	Wibson 
}

\maketitle

\begin{abstract}

We are tackling the problem of trading real-world private information using only cryptographic protocols and a public blockchain to guarantee honest transactions. In this project, we consider three types of agents ---buyers, sellers and notaries--- interacting in a decentralized privacy-preserving data marketplace (\dPDM) such as the Wibson data marketplace. This framework offers infrastructure and financial incentives for individuals to securely sell personal information while preserving personal privacy. Here we provide an efficient cryptographic primitive for the secure exchange of data in a \dPDM, which occurs as an atomic operation wherein the data buyer gets access to the data and the data seller gets paid simultaneously. 

\end{abstract}

%---------------------------------------

%---------------------------------------
\section{Preliminaries}
%---------------------------------------

In this paper, we are interested in the problem of trading real-world private information using only cryptographic protocols and a public blockchain to guarantee honest transactions.
Private information in this context refers to attributes or events associated with a single individual (person or organization). %, or its relationship with other individuals. 

In \cite{Travizano2018Wibson} the authors introduced the notion of a decentralized Privacy-Preserving Data Marketplace (\dPDM).
A decentralized Data Marketplace (\dDM) is a Data Marketplace (\DM) with no central authority, no central data repository and no central funds repository.
Additionally, a {\dPDM}  allows users to sell private information, while providing them privacy guarantees such as:
\begin{itemize}
\item Participants anonymity: the identities of the Sellers and Buyers are not revealed without their consent. In particular, the identity of the Data Seller is not revealed to the Data Buyer, without the consent of the Data Seller.
\item Transparency over Data usage: the Data Seller always has visibility on how his Data is used by the Buyer.
\end{itemize}

Here we consider three types of agents interacting in a {\dPDM} both privately (through end-to-end communications) and publicly through a permissionless blockchain:
\begin{description}
\item[Data Seller:]
The Seller $\fraS$ is the owner and subject of the private information that will be traded. He decides when and if his data is sold.

\item[Data Buyer:]
The Buyer $\fraB$ is interested in acquiring information from Sellers, provided that the information meets the Buyer's quality requirements.

We expect Data Buyers to be organizations that will receive information from willing and actively participating individuals to train Data Science algorithms and models, with the benefit of knowing that the personal information should be accurate and current.

\item[Notary:]
The Notary $\fraN$ has the means to validate information associated with Sellers.  He is trusted by Buyers to certify the precision and quality of the data traded.   
The Notary is the only public player with a formal track record and a public reputation. 

\end{description}

The information traded is collected mainly outside of the blockchain. The Notary will typically access the data as part of its business operations with the Sellers. The Buyer understands the value associated with the privileged position of the Notary and knows about the incentives aligned with its reputation.

We illustrate the three market roles with an example.
Suppose that the Seller is a client of a Bank, who offers on the market his (anonymized) credit card transactions. The Buyer can be any entity requiring transactional data to train its Machine Learning models. 
In this example, the Bank is the ideal Notary since:
\begin{itemize}
\item The Bank can verify that the Seller is actually a client of the Bank, by requiring the Seller to provide information that authenticates him/her.

\item The Bank can act as a Notary in case of conflict, and verify whether the information of credit card transactions sent by the Seller to the Buyer is valid and trustworthy (in particular, by comparing with the Bank's own records of the client's credit card transactions).

\end{itemize}

%---------------------------------------
\section{Problem Statement and Related Work}
%---------------------------------------

We assume that the Data Seller $\fraS$ and Data Buyer $\fraB$ are participating in a Data Exchange protocol, such as the Wibson protocol~\cite{Travizano2018Wibson}, and have already completed the following steps:
\begin{itemize}
\item The Buyer $\fraB$ has verified that the Seller $\fraS$ belongs to the Buyer’s audience of interest.
\item The Data requested is available.
\item Buyer $\fraB$ and Seller $\fraS$ have agreed on a price that is acceptable to both parties.
\end{itemize}

They are now faced with the following challenges:
\begin{enumerate}[label={(\arabic*)}]
\item If the Seller releases the information first, the Buyer may decide not to pay.
\item If the Buyer pays first, the Seller may decide not to reveal the information.
\item The Seller may reveal incomplete or false information.
\end{enumerate}

Challenges (1) and (2) are known as the problem of \textbf{fair exchange}, which has been studied for decades.
Study~\cite{cleve1986limits} showed that fairness
is unachievable without the aid of a trusted third party.
However, the blockchain can fill the role of the trusted party, and essentially eliminates the trust problem.

Zero Knowledge Contingent Payment (ZKCP) protocols have been proposed, which allow the fair exchange of goods and payments over the Bitcoin network~\cite{campanelli2017zero}.
The ZKCP protocols require the execution of a zero-knowledge proof in order to work.
Interesting applications of the zero-knowledge proof of binding provided by ZKCP include the query to a database of passwords and hashes~\cite{zkcp2018}, a problem tackled from a Private Information Retrieval perspective in~\cite{calvo2012oblivious}.

When ZKCP was first introduced in 2011 it was only theoretical as there were no known efficient zero-knowledge protocols that could be used for the proofs at that time. Since then, advances have been made and there are now general-purpose Succinct Non-Interactive Arguments of Knowledge (zk-SNARK) protocols that allow the  implementation of the necessary proofs~\cite{sasson2014zerocash,ben2018scalable}.

However, the zk-SNARK protocols are expensive in terms of computational and transactional cost required to execute them on a blockchain such as Ethereum.
Here we propose a solution that leverages the trust that the Buyer places in the Notary to solve Challenge (3) (ensure data quality) as well as Challenges (1) and (2) (fair exchange) in a very efficient way.

%---------------------------------------
\section{Secure Exchange of Digital Goods}
%---------------------------------------

%---------------------------------------
\subsection{Protocol SEDG1}
%---------------------------------------

The first contribution of this paper is the protocol described in Table~\ref{tab:sedg1}, which enables a ``Secure Exchange of Digital Goods'' (SEDG) by leveraging the Buyer's trust in the Notary in order to solve the challenges previously mentioned.

\begin{table}[ht]	
\caption{Secure Exchange of Digital Goods (protocol 1)}
\label{tab:sedg1}
\medskip
\begin{center}
  \begin{tabular}[t]{ l l l }
    \toprule
Notary $\fraN$	& Seller $\fraS$ 	& Buyer $\fraB$ \\ 
\midrule
$ k = \mbox{Random}() $ & & \\
$ C = E_k(\mbox{Data}_{\fraS}) $ & & \\
$ h_1 = H(C) $  & & \\
$ h_2 = H(k) $ & & \\
$ \sigma = \mbox{Sign}_{Notary}( h_1 || h_2 || \fraS_{id}) $		& & \\ 
\midrule
$ \mbox{Send}_{Seller} ( k || C || \sigma ) $		& & \\ 
\midrule
	& $ \mbox{Send}_{Buyer} ( \sigma || C || h_2 ) $	 & \\
\midrule
	& & $ \mbox{Verify} (\sigma) $ \\ 
	& & Check $ H(C) = h_1$? \\
\midrule
	& & $ T(x) := Pay(Seller) $ \\
	& &  \; \; if $ H(x) = h_2 $ \\
	& & $ \mbox{Publish}(T) $ \\
\midrule
	& $ \mbox{Publish}_{T} (k) $	& \\
    \bottomrule
  \end{tabular}
\end{center}
\end{table}

There is an initial \textbf{setup phase}, during which the Notary $\fraN$ receives the Seller's  associated information ($\mbox{Data}_{\fraS}$) and generates a certificate by performing the following steps:
\begin{enumerate}
\item Notary verifies the information $\mbox{Data}_{\fraS}$ received from the Seller $\fraS$. $\mbox{Data}_{\fraS}$ contains the actual Data as well as the Seller's identification $\fraS_{id}$ and meta-information about the type of data.
\item Notary generates a random key $k$ and uses $k$ to create an encrypted version of the data ($C$).
\item Notary generates commitments for the key and ciphertext ($h_1$ and $h_2$), by computing a secure hash $H$ of the key and ciphertext.
\item Notary signs a string obtained by concatenating $h_1$, $h_2$ and the Seller's identification $\fraS_{id}$.
\item Finally the Notary sends to the Seller $\fraS$ a certificate containing
the random key $k$, the encrypted data $C$ and the signature $\sigma$.
\end{enumerate}

During the \textbf{transaction phase} (when Buyer and Seller actually perform the Exchange of Digital Goods), they follow these steps:
\begin{enumerate}
\item Seller $\fraS$ sends to the Buyer the signature $\sigma$, the encrypted data $C$ and $h_2$.
\item The Buyer can verify that the signature $\sigma$ is correct.
\item The Buyer can check that it has a proper encryption of the data by verifying the opening of the commitment for $C$, that is by verifying whether $H(C)$ is equal to $h_1$.
\item If everything is correct, the Buyer publishes a transaction on the blockchain that will pay the Seller if the key is revealed.
The payment is executed only if the Seller exhibits an $x$ such that $H(x) = h_2$.
\item The Seller closes the transaction $\fraT$ by publishing $k$, effectively opening the commitment $h_2$.
\end{enumerate}

Note that the Notary generates the Seller's certificate before (or independently from) the Data request by the Buyer. The Notary does not play any part in the protocol during the transaction phase.

After the transaction is completed, the Buyer $\fraB$ uses the encryption key $k$ to gain access to the Seller's data. This mechanism, wherein certain content is maintained private until a particular event (the publication of $k$) occurs, is reminiscent of the family of cryptographic primitives called Secure Triggers~\cite{futoransky2006foundations}.

%---------------------------------------
\subsection{Protocol SEDG2}
%---------------------------------------

In the Wibson Data Marketplace~\cite{Travizano2018Wibson}, the Notary $\fraN$ is paid for his services in respect to a Data Transaction $\fraT$ by receiving part of the tokens paid by the Buyer.

\begin{table}[ht]	
\caption{Secure Exchange of Digital Goods (protocol 2)}
\label{tab:sedg2}
\medskip
\begin{center}
  \begin{tabular}[t]{ l l l }
    \toprule
Notary $\fraN$	& Seller $\fraS$ 	& Buyer $\fraB$ \\ 
\midrule
$ k = \mbox{Random}() $ 		& & \\
$ C = E_k (\mbox{Data}_{\fraS}) $		& & \\
$ h_1 = H(C) $		& & \\
$ h_2 = H (k || \fraN_{id}) $	& & \\
$ \sigma = \mbox{Sign}_{Notary} (h_1 ||h_2 || \fraS_{id}) $	& & \\ \midrule	
$ \mbox{Send}_{Seller} ( k || C || \sigma ) $		& & \\ 
\midrule
	& $ \mbox{Send}_{Buyer} ( \sigma || C || h_2 ) $	 & \\ \midrule
	& & $ \mbox{Verify} (\sigma) $ \\
	& & Check $ H(C) = h_1$? \\ \midrule
	& & $ T(x,n) :=  \mbox{Pay} (Seller, n) $ \\
	& & \; \; if $ H( x || n ) = h_2 $ \\
	& & $ \mbox{Publish} (T) $ \\ \midrule
	& $ \mbox{Publish}_{T} (k, \fraN_{id}) $	& \\
    \bottomrule
  \end{tabular}
\end{center}
\end{table}

In this section we describe a variation of the SEDG protocol, that allows the Notary to be paid for his services simultaneously with the Data Seller $\fraS$. 
The solution is described in Table~\ref{tab:sedg2}.

The difference with the protocol SEDG1 is to include the Notary identification $\fraN_{id}$ as part of the key commitment to guarantee the payment:
\begin{itemize}
\item During the \textbf{setup phase}, $h_2$ is computed as the hash of the random key $k$ and the Notary identification $\fraN_{id}$. 
\item During the \textbf{transaction phase}, in order to close the transaction $\fraT$, the Seller $\fraS$ has to publish both the
encryption key $k$ and the identification $\fraN_{id}$.

\end{itemize}

%---------------------------------------
\subsection{Protocol SEDG3}
%---------------------------------------

Finally, we tackle the problem of hiding the transaction $\fraT$ from the Notary $\fraN$. In this scenario, after receiving the certification from the Notary, the Data Seller $\fraS$ wants to be able to use the certification without the Notary learning that the certification is being used.

\begin{table}[ht]	
\caption{Secure Exchange of Digital Goods (protocol 3)}
\label{tab:sedg3}
\medskip
\begin{center}
  \begin{tabular}[t]{ l l l }
    \toprule
Notary $\fraN$	& Seller $\fraS$ 	& Buyer $\fraB$ \\ 
\midrule
$ k = \mbox{Random}() $		& & \\
$ C = E_k (\mbox{Data}_{\fraS}) $		& & \\
$ h_1 = H(C) $		& & \\
$ h_2 \equiv G^k \mod p $		& & \\
$ \sigma = \mbox{Sign}_{Notary} ( h_1 || h_2 || \fraS_{id}) $		&  & \\ 
\midrule
$ \mbox{Send}_{Seller} ( k || C || \sigma ) $		&  & \\ \midrule
	& $ \mbox{Send}_{Buyer} ( \sigma || C || h_2 ) $	& \\ \midrule
	& & $ \mbox{Verify} (\sigma) $ \\
	& & Check $ H(C) = h_1$? \\
	& & $ r = \mbox{Random}() $ \\
	& & $ \mbox{Send}_{Seller}(r) $  \\ \midrule
	& & $ T(x) := \mbox{Pay}(Seller) $ \\
	& & \; \; if $G^x \equiv {h_2}^r \mod p $ \\ \midrule
	& & $ \mbox{Publish}(T) $ \\ \midrule
	& $ \mbox{Publish}_{T} ( k \cdot r ) $ 	 & \\
    \bottomrule
  \end{tabular}
\end{center}
\end{table} 

The following protocol uses a blinded commitment~\cite{pedersen1991non} to hide the transaction from the Notary. It uses a public generator $G$ to create the commitments. A simple version based on the discrete-log assumption in described in Table~\ref{tab:sedg3}.

The difference with the protocol SEDG1 is that it uses a finite field generator $G \in \Z_p$ for the commitment $h_2$:
\begin{itemize}
\item During the \textbf{setup phase}, $h_2$ is computed as a power of $G$ to the random key $k$ in the finite field $\Z_p$.
\item During the \textbf{transaction phase}, the Buyer chooses an additional random number $r$ and sends it to the Seller.
\item To close the transaction $\fraT$, the Seller $\fraS$ has to publish the product $k \cdot r$, thus effectively hiding the original key $k$ from the Notary.
	If the transaction is correct, when $\fraS$ publishes $x$, it holds that
	 $$G^{x} \equiv G^{(k \cdot r)} \equiv {\left( G^k \right)} ^ {r} \equiv {h_2}^r \mod p .$$
	
\end{itemize}

As a result of the protocol SEDG3, the Notary $\fraN$ will not be able to identify the published transaction $\fraT$ as belonging to Seller $\fraS$.

%---------------------------------------
\section{Conclusion}
%---------------------------------------

Here we proposed a solution to the problem of trading real-world private information using only cryptographic protocols and a public blockchain to guarantee the fairness of transactions.
We described a protocol that we call ``Secure Exchange of Digital Goods'' (SEDG) between a Data Buyer $\fraB$ and a Data Seller $\fraS$.
The protocol relies on a trusted third party $\fraN$, which also plays the role of Notary in the context of a decentralized Privacy-Preserving Data Marketplace (\dPDM) such as the Wibson Marketplace~\cite{Travizano2018Wibson}.

This protocol converts the Exchange of Data into an atomic transaction where two things happen simultaneously:
\begin{itemize}
\item The Buyer $\fraB$ gets access to the Data, by learning the key that enables him to decrypt $C$ (previously received encrypted data).
\item The Seller $\fraS$ gets paid for his Data by revealing the key.
\end{itemize}

We also presented two variations of the base protocol:
\begin{itemize}
\item  In SEDG2, the Notary gets paid for his services at the very same time that the Seller gets paid for his data.
\item In SEDG3, the Data Transaction is hidden from the Notary $\fraN$ (who generated the certificates used during the transaction).
\end{itemize}

\section*{Acknowledgements}

The authors thank Nicolás Ayala and Martin Manelli 
for their work on the implementation of these solutions.

%----------------------------------------------------
\bibliographystyle{unsrt}
\bibliography{../wibson}
%----------------------------------------------------

\end{document}